\documentclass{article}
\usepackage[utf8]{inputenc}
\usepackage[margin = 3cm]{geometry}
\usepackage{graphicx}
\usepackage{amssymb}
\usepackage[colorlinks = true,
            linkcolor = blue,
            urlcolor  = blue,
            citecolor = blue,
            anchorcolor = blue]{hyperref}
\usepackage{caption}
\usepackage{subcaption}
\usepackage{textcomp}
\usepackage{cite}
\usepackage{soul}
\usepackage{authblk} 

\begin{document}

\author[1]{Elia~Bertoldo\thanks{Corresponding author: ebertoldo@ifae.es}}
\author[2]{Victor~P\'erez~S\'anchez}
\author[3]{Maria~Mart\'inez}
\author[1,4]{Manel~Mart\'inez}
\author[5]{Hawraa Khalife}
\author[1,4]{Pol~Forn-D\'iaz}
\affil[1]{Institut de F\'isica d’Altes Energies, The Barcelona Institute of Science and Technology, Bellaterra 08193, Spain}
\affil[2]{Laboratorio Subterráneo de Canfranc, 22880 Canfranc-Estación, Spain}
\affil[3]{Centro de Astropartículas y Física de Altas Energías, Universidad de Zaragoza, 50009 Zaragoza, Spain}
\affil[4]{Qilimanjaro Quantum Tech, Barcelona 08007, Spain}
\affil[5]{IRFU, CEA, Université Paris-Saclay, 91191 Saclay, France}

\setcounter{Maxaffil}{0}
\renewcommand\Affilfont{\itshape\small}

\title{Cosmic muon flux attenuation methods for superconducting qubit experiments}

\maketitle

\begin{abstract}
    We propose and demonstrate two practical mitigation methods to attenuate the cosmic muon flux, compatible with experiments involving superconducting qubits: shallow underground sites and device orientation. Using a specifically-built cosmic muon detector, we identify  underground sites, widely present in urban environments, where significant attenuation of cosmic muon flux, up to a factor 35 for 100-meter depths, can be attained. Furthermore, we employ two germanium wafers in an above-ground laboratory, each equipped with a particle sensor, to show how the orientation of a chip with respect to the sky affects the amount and type of energy deposited on the substrate by ionizing radiation. We observe that the horizontal detector sees more counts at lower energy, while the vertical one is impacted by more particles at higher energy.  The methods here described proposed ways to directly understand and reduce the effects of cosmic rays on qubits by attenuating the source of this type of decoherence, complementing existing on-chip mitigation strategies. We expect that both on-chip and off-chip methods combined will become ubiquitous in quantum technologies based on superconducting qubit circuits.
\end{abstract}
\section*{Introduction}

Superconducting qubit circuits are sensitive probes to many types of fluctuations present in their environment. These fluctuations place a limit to the qubit coherence time. The coherence times already achieved render qubits sensitive to very subtle environmental events, such as stochastic ionizing radiation impacts coming either from background radioactivity~\cite{vepsalaainen2020} or from cosmic muons~\cite{cardani2021}. While environmental radioactivity could be screened using radiopure materials as well as a dedicated shielding, as is routinely implemented in low-background experiments~\cite{Heusser:1995wd}, cosmic muons cannot be screened. As such, the only method to drastically reduce the muon flux nowadays is to move the experiment to deep underground laboratories.

Ionizing radiation events are currently not the dominant source of superconducting qubit decoherence~\cite{Gordon2022}. Nevertheless, increasing evidence points to this radiation being responsible for qubit device stability and performance~\cite{vepsalaainen2020}, through its effects lingering on the qubit substrate in the form of phonons~\cite{mcewen2022}, quasiparticles~\cite{wilen2021}, and the scrambling of two-level defects~\cite{Thorbeck:2022yzs}. Therefore, techniques to mitigate the impact of ionizing radiation events are already necessary today. While on-chip mitigation strategies, such as phonon~\cite{iaia2022} or quasiparticle traps~\cite{henriques2019, Riwar2016, bargerbos2022}, are being developed, it is also necessary to understand the physics behind ionizing radiation, particularly the cosmic muon events, in order to find ways to attenuate its flux. In this work, we propose two such methods. The first method is to identify shallow underground sites at accessible areas such as urban environments, considering the attenuation factors of muon flux already characterized at such moderate depths~\cite{Szucs:2019awp, Bogdanova:2006ex, PNNL, TUM}, which may already set no limitations to qubit coherence. The second method studies device-orientation dependence with respect to the zenith angle, since muons reaching the Earth's surface have a strong angular dependence due to their absorption through the atmosphere. Specifically, for the average energy of muons reaching the Earth's surface ($\sim$4 GeV) a cosine-square dependence is expected.

We perform a muon flux characterization at different depths using a home-built muon detector and verify the data collected with tailored Geant4 simulations. We also provide a detailed description of the detector so that it can be reproduced. \\
In parallel, we measured the amount of energy deposition by ionizing radiation at cryogenic temperatures for two identical germanium wafers: this experiment showed how the orientation of the chip with respect to the sky can strongly change the energy deposition within the substrate, mainly due to the strong angular dependence of cosmic rays reaching the device.

We believe that superconducting qubit technology developers need to start bringing their attention to ionizing radiation, and cosmic muons in particular, in order to push even further the quality and stability of the devices in order to develop universal quantum computers and quantum sensors for the detection of, e.g., rare physics events.

This work is organized as follows: in Section~\ref{sec1} we give an introduction of the physics of cosmic rays and its connection with qubits. In Section~\ref{sec2}, we describe our specific muon detector and present its characterization. Section~\ref{sec3} presents our main results of the dependence of cosmic muon flux on the rock depth for shallow sites along with the Geant4 simulations developed to cross-check the data collected. Section~\ref{sec4} presents the difference in energy spectra recorded in two germanium wafers, one oriented parallel with respect to the sky (zenith angle of 90 degrees), the other orthogonal with respect to the sky (zenith angle of 0 degrees). Finally, we provide conclusions and future perspectives in Section~\ref{conclusions}.

\section{Introduction to Cosmic muon physics for qubits}
\label{sec1}
\subsection{Cosmic rays}
\label{sec1.1}

There are multiple particles of cosmic origin that reach Earth. These particles can be divided in two broad categories: charged and neutral. Cosmic rays is a terminology that usually defines charged particles that originate outside the solar system~\cite{Schroder:2016hrv, Letessier-Selvon:2011sak}. The neutral particles reaching Earth are neutrinos and photons, with neutrinos generally not interacting with the atmosphere. This differentiates neutrinos from charged particles and photons, which normally interact with the atmosphere and generate air showers of secondary cosmic rays.
Primary cosmic rays are generated within our galaxy and by extragalactic sources. Primary cosmic rays which reach our atmosphere are almost exclusively protons ($\sim$90\%) and alpha particles ($\sim$9\%), with small percentages of heavier nuclei and electrons. An even tinier percentage is composed by antimatter particles in the form of positrons and antiprotons. The energy of this primary cosmic rays spans several orders of magnitude, with maximum energies exceeding 10$^{20}$~eV~\cite{Letessier-Selvon:2011sak}.

In this work, we will focus on the type of cosmic rays that have the ability to reach the surface and the underground laboratories on Earth. The particles reaching the surface are mostly the products of the interactions between primary cosmic rays and the atmosphere, while the particles reaching underground labs can also be the product of the interactions of secondary cosmic rays with the surrounding rocks and materials.

\subsubsection{Cosmic rays at sea level}
\label{sec1.1.1}

When primary cosmic rays enter the atmosphere they start interacting with atoms and molecules, producing a cascade of particles, often referred to as secondary cosmic rays (which include photons, protons, alpha particles, pions, muons, electrons, neutrinos, and neutrons). At this point, the picture starts to become more complicated, with numerous unstable and fast-decaying particles created, and a variety of more stable particles that keep advancing towards Earth's surface. In general, many photons are produced in this interaction which produce electromagnetic showers~\cite{Schroder:2016hrv}. The rest of particles might also eventually decay into photons and electrons, feeding the electromagnetic showers. If hadrons survive or are newly produced, the shower continues its development. However, after a certain number of generations, the shower starts to fade, with fewer new particles produced than the ones absorbed in the interaction with the atmosphere~\cite{Schroder:2016hrv}. 
Muons and neutrinos created in these processes do not contribute to the shower development and approximately continue freely their path towards our planet. As a result, the vast majority of particles reaching the sea level are neutrinos and muons, while electrons, positrons, photons, neutrons, and protons compose a small fraction of the total flux~\cite{Workman:2022ynf}.

Muons, which are the main focus of this work, are elementary particles of the lepton family with an electric charge of -1e, spin 1/2, and a 105.66 MeV/c$^2$ mass~\cite{Workman:2022ynf}. They are also unstable, with a mean lifetime of 2.2~$\mu$s~\cite{Workman:2022ynf}. At sea level, muons exhibit a rate of approximately 1 count per minute per cm$^2$ of horizontal surface and a mean energy around 4~GeV~\cite{Rocca18}.

\subsubsection{Cosmic rays underground}
\label{sec1.1.2}
In underground laboratories, the muon flux decreases almost exponentially with the thickness of the rock overburden above the laboratory~\cite{Bettini:2012fw} (see Section~\ref{sec3}). Normally, to compare the depth of different laboratories, since the rock composition varies, the overburden is converted in meter water equivalent (m.w.e.)~\footnote{The conversion between m.w.e. and real meters depends on the rock composition, with typical values in the range of 2.4-3~m.w.e./m.}.

At shallow depths ($\sim10$~m.w.e.), only muons, protons, and neutrons constitute significant sources of cosmic ray background. In particular, interactions caused by neutrons are extremely difficult to suppress. These neutrons are usually a product of proton and muon spallation in the rocks and materials surrounding the experimental setup. Designing effective shields against neutrons is particularly cumbersome, due to their ability to cross large distances.

Thick lead shields, ordinarily used to effectively suppress environmental radioactivity, are the perfect example of the challenges that can arise in above-ground laboratories when trying to reduce the ionizing radiation reaching the experimental volume. In fact, ordinary lead shields are also contaminated with radioisotopes, the most concerning of which is $^{210}$Pb~\cite{Beeman:2022wun}. To avoid this problem, archaeological lead with a low concentration of radionuclides can be employed~\cite{CDMS:2002moo, ARNABOLDI2004}. Lead can be effective in reducing environmental radioactivity, but it can also be a source of additional radiation: due to its high density, lead is a perfect medium for spallation events, which create neutrons in the medium. To reduce the flux of neutrons generated by spallation in lead and other materials surrounding the cryostat, usually a polyethylene shield is interposed between the cryostat and the lead shield~\cite{CDMS:2002moo}. It should be noted that the polyethylene shield thickness should be carefully selected, in order not to have a relevant production of spallation on polyethylene itself~\cite{CDMS:2002moo}.

At the surface level, almost all spallation events are due to proton interactions, but below 15~m.w.e., due to the strong attenuation of the proton flux, spallation induced by muons becomes dominant~\cite{PNNL}. As such, at 30~m.w.e. the muon flux is reduced by approximately a factor 6, but the neutron flux is reduced by a factor of almost 100~\cite{Formaggio:2004ge, PNNL}.
In fact, several shallow laboratories exist around the world where muon flux screening factors of 5-10 are attained without the need of very large depths~\cite{PNNL}.

These shallow laboratories offer various advantages: they reduce the amount of direct muon interactions with the sample, they reduce the number of fast neutrons generated by spallation events caused by cosmic protons and muons, and they also reduce the cosmogenic activation of materials constituting or surrounding the experimental setup~\cite{Cebrian:2017oft, Zhang:2016rlz}. In addition, and contrary to deep underground sites, shallow laboratories present far less difficulties to be established at accessible areas such as urban environments.

At significant depths ($>1$~km.w.e.), only muons and neutrinos can cause events in the experimental volume, both by direct hits or by inducing tertiary fluxes of particles due to interactions with surrounding rocks and materials~\cite{Workman:2022ynf}.



The only viable solution to drastically reduce the cosmic radiation background due to cosmic muons is to perform experiments in deep underground laboratories. Several laboratories of this type exist around the world and they generally host experiments focused on astroparticle physics~\footnote{
The current list of operative deep underground laboratories in Europe includes the Baksan Neutrino Observatory (BNO) in the Russian Federation~\cite{Kuzminov:2012fv}, Boulby in the United Kingdom~\cite{Boulby}, the Canfranc Laboratory (LSC) in Spain~\cite{Canfranc}, the Centre for Underground Physics in Pyhäsalmi (CUPP) in Finland~\cite{CUPP}, the Gran Sasso Laboratory (LNGS) in Italy~\cite{LNGS}, and the Modane Laboratory (LSM) in France~\cite{Modane}. In Asia, we can find the China Jinping Underground Laboratory (CJPL) in China~\cite{CJPL}, the Kamioka Observatory in Japan~\cite{Kamioka}, the Yangyang Underground Laboratory~\cite{Yoon_2021} and Yemilab~\cite{Park_2024} in Korea. In Australia there is the Stawell Underground Physics Laboratory (SUPL)~\cite{Urquijo:2016dxd}. Finally, in North America, one counts the Sanford Underground Research Facility (SURF)~\cite{SURF}, the Soudan Underground Laboratory~\cite{Soudan}, the Waste Isolation Pilot Plant (WIPP)~\cite{WIPP}, and the Kimballton Underground Research Facility (KURF)~\cite{KURF} in the USA, and SNOLAB in Canada~\cite{SNOLAB}.}.
The deepest operating underground lab is CJPL in China while the biggest in size is LNGS in Italy~\cite{pic3}.

\subsection{Qubit decoherence due to ionizing radiation}
\label{sec1.4}

In recent years, evidence has been mounting on the detrimental effect of ionizing radiation on superconducting qubits~\cite{martinis2021}, particularly from cosmic rays. At least two mechanisms have been identified so far originated from particle impacts on the bulk of the device substrate, indirectly impacting qubit performance.

The first mechanisms corresponds to an impinging high-energy particle or photon producing highly energetic electrons in the substrate which excites electron-hole pairs, themselves producing phonons upon recombination. These highly energetic phonons propagate quasi-diffusively and then promptly decay to lower-energy states that travel ballistically throughout the chip. Some of these phonons eventually scatter at the boundary with the superconductor leading to the breaking of Cooper pairs and thus to the creation of quasiparticles which are a known source of qubit decoherence~\cite{catelani2011}.

The second mechanism has been recently identified to be a scrambling effect on the two-level systems~\cite{Thorbeck:2022yzs} (TLSs), believed to reside mostly at the superconductor-insulator interface of the qubit circuit~\cite{lisenfeld2019}. These TSLs display a very broad frequency spectrum, often coming into resonance with qubits they couple to, leading to either fluctuations in the relaxation time ($T_1$)~\cite{Thorbeck:2022yzs} or gaps in the energy spectrum~\cite{neeley2008}. The scrambling effect causes shifts in the frequency of TLSs and thus it is responsible for both frequency noise in qubits as well as fluctuations in $T_1$ and the decoherence time ($T_2$)~\cite{Klimov2018}.

Both mechanisms are particularly harmful as they cause correlated noise~\cite{mcewen2022} which cannot be accounted for in currently explored quantum error correcting methods such as the surface code~\cite{mariantoni2012}. Another possible mechanism not currently studied are direct hits on the superconducting material, similar to the detection mechanism of single photons in superconducting nanowire photon detectors~\cite{dorenbos2011}. The deleterious effect of cosmic rays was also demonstrated on superconducting resonators placed in the deep underground laboratory in Gran Sasso~\cite{cardani2021}. Other evidence of the impact of ionizing radiation has been demonstrated on multi-qubit devices, where correlated charge fluctuating events between neighboring qubits were observed~\cite{wilen2021}, and in the stability of fluxons (the magnetic flux quanta) in superconducting loops~\cite{Gusenkova_2022}.

The first direct evidence of the impact of ionizing radiation on a superconducting qubit used a radioactive source installed directly inside of a dilution refrigerator mixing chamber stage~\cite{vepsalaainen2020}. A transmon qubit placed in the vicinity of the source experimented a drastic decrease of its lifetime due to the activity of the source. Based on those measurements, an upper bound of the transmon qubit lifetime due to background ionizing radiation of more than 3 milliseconds was established, with an estimated 40\% of the external background events corresponding to cosmic rays. Other works using fluxonium qubits have reached millisecond-range lifetimes~\cite{pop2014} and coherences~\cite{Somoroff:2021elj}, possibly limited by quasiparticles (QPs) partially generated by ionizing radiation as well as by stray microwave fields~\cite{Pan2022}.

Assuming one is able to eliminate all background radioactivity by proper shielding inside and outside the cryostat, only cosmic rays will impact qubit coherence. A decrease of the cosmic muon flux would lead to a reduction of the steady-state amount of excess QPs, given the reduction of energetic phonons in the substrate. The relation between high energy impacts and QP density is complex and depends on several material properties. In addition, a reduction of cosmic muon flux would also reduce the rate of TLS fluctuating frequency, leading to more stable qubit frequencies and coherence times.

Environmental radioactivity from neighboring sources inside the dilution refrigerator are progressively being identified~\cite{Cardani:2022blq}. Attention will be placed in future experimental setups and laboratory spaces towards more radiopure systems, such as those employed in complex, low-background experiments~\cite{ARNABOLDI2004}. Eventually, the only significant remaining source of ionizing radiation will be the cosmic rays, which, as stated in the previous section, cannot be easily screened above ground. Therefore, mitigation strategies are needed to suppress the effects of cosmic rays. In this work, we propose two methods: the screening obtained in shallow underground sites (Section~\ref{sec3}) and the specific device orientation (Section~\ref{sec4}). Other initiatives in the field have been addressing the same problem with direct on-chip mitigation strategies, such as lower-gap superconducting quasiparticle traps~\cite{henriques2019}, phonon traps~\cite{iaia2022}, or suspending the device on a membrane~\cite{karatsu2019}. All strategies have evidenced a decrease on the impact of ionizing events on the qubits and/or resonators employed as sensors. Very likely, the final solution will become a combination of both external as well as internal mitigation methods. Finally, a novel unique strategy proposes to use a cryogenic veto system compatible with the operation of superconducting qubit devices~\cite{10313842}.

\section{Compact muon detector}
\label{sec2}
In this section, we present a compact and portable cosmic muon detector that we purposely built to characterize the flux of muons at shallow underground sites of different depths.

\subsection{Detection principle}
The detector we built is based on two superimposed plastic scintillators coupled to photo multiplier tubes (PMTs) and readout in coincidence for the detection of cosmic muons. The plastic scintillator used emits fast pulses of light when radiation crosses it, with a spectrum and speed well matched to the spectral sensitivity and time response of the PMT.

In principle, these detector setups are meant to be used to detect any type of radiation strong enough to create a detectable signal. Therefore, in order to discriminate muons, since they are penetrating particles, we have chosen to use two identical detectors physically superimposed to each other and connected to a coincidence circuit. A thick (1.7~cm) metal shielding made of copper between both detectors helps in increasing the filtering for particles other than muons.

\subsection{Description of the detector}

The scintillator-PMT setup comes already pre-assembled as a commercial detector unit (SCIONIX model R30*20B250/1.1-E3-P-X) which includes an Ej200 plastic scintillation detector of dimensions ($20\times30\times250)~$mm$^3$ wrapped in reflector and surrounded by light tight vinyl coupled to 30~mm diameter R6094 PMT equipped with a voltage divider and counting electronics. The first 1.5~mm wrapping layer of this detector is not sensitive to radiation.

\begin{figure}
\centering
\includegraphics[width=0.8\textwidth]{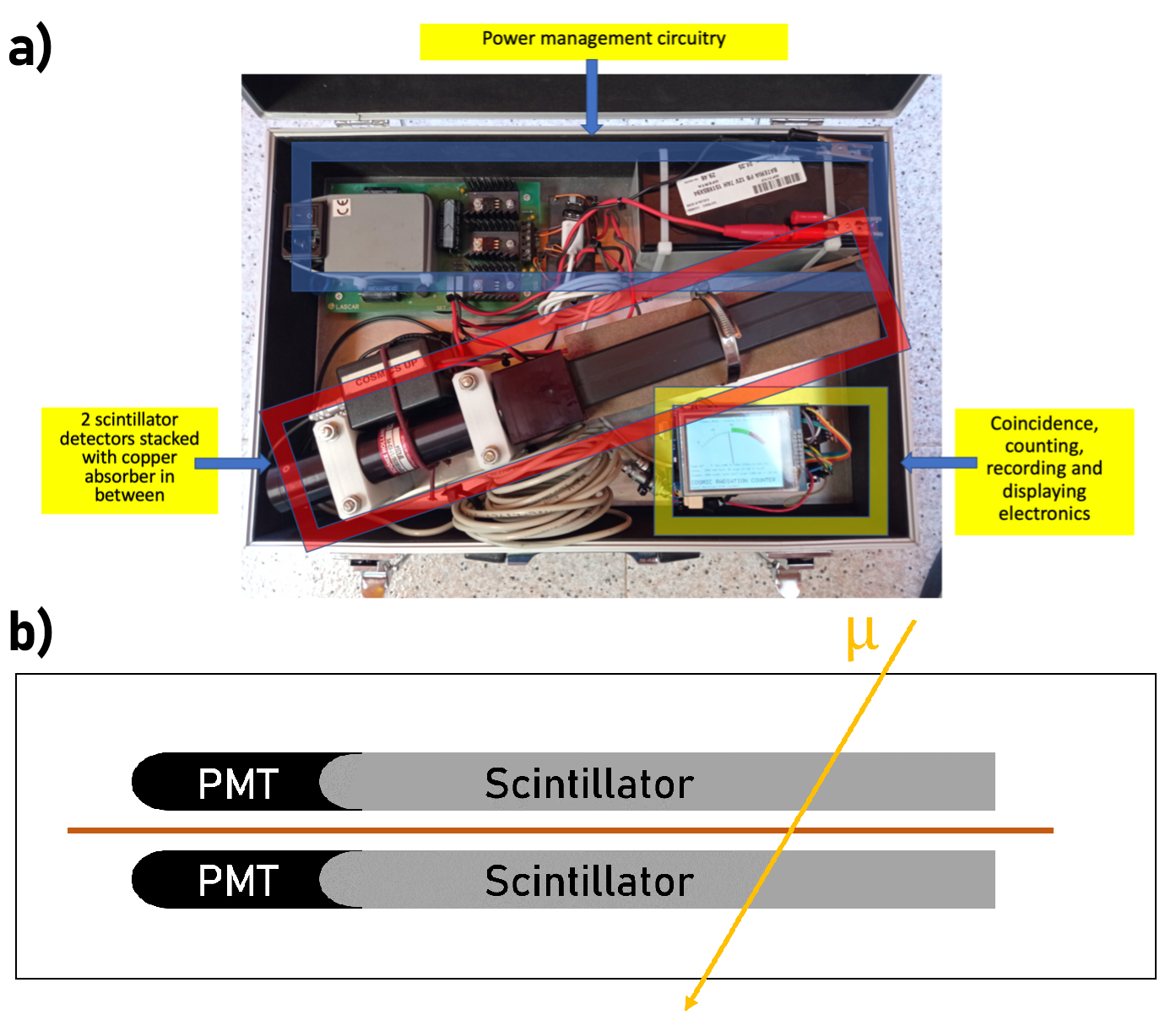}
\caption{a) The muon detector inside the suitcase. The upper part contains the power circuitry including the network inlet, the power supply and the low-voltage regulators, the HV generators, and the lead battery for standalone operation without power connection. In the center of the case we positioned the two scintillator detectors set stacked with a copper absorber in between, the bottom-right contains the coincidence, counting, recording, and displaying circuitry. b) The schematic of the detector inside the suitcase, seen from the side: two scintillators (in grey) coupled to a PMT each (in black). A copper plate is located between the two detectors to absorb particles other than muons. In yellow, a typical muon interaction leading to a count, with the muon crossing both scintillators.  \label{fig:inside}}
\end{figure}

The detector has been built in a very compact setup for easy transportation and fits into a standard size ``executive" suitcase (see Fig.~\ref{fig:inside}). It is powered by a 12~V lead battery or alternatively connected to the electrical network. In this second case, the battery is recharged in parallel. From either the 12~V from the battery or the AC power supply delivering 12~V output, a power management circuitry delivers all the needed voltages (see Fig.~\ref{fig:accidental}).
Using DC-DC converters, from these 12~V the setup generates the high voltages needed to power the PMTs (700~V) and, using linear regulators, the 5~V needed to power the coincidence circuit and the microcontroller that takes care of the DAQ, display and data saving.

Each scintillator plus PMT detector setup provides already a TTL digital discriminated output, that is used as input for the coincidence electronics. The length of the digital output is of 50~ns while the signal-overlap time requested for detecting a coincidence (resolving time) is of about 20~$\mu$s. The size of this coincidence window is limited by the internal clock of the microcontroller.
Each single detector signal, as well as their coincidences, are input to an Arduino Mega microcontroller that, using interrupts, counts them and computes their rate and their statistical uncertainty. The microcontroller also measures the setup temperature, humidity and atmospheric pressure and displays all the quantities mentioned into a large display 
 while recording all data into an SD card in easily readable text format for later analysis.
The suitcase can be easily carried and installed in any place, while the battery provides an autonomy of a few days.

{\subsection{Detector characterization}
\label{sec2.3}

In order to assess the accidental coincidence rate $N_a$, one can use the single rate measurements $N_i$ from each detector together with the coincidence resolving time $\tau$ in a simple probabilistic calculation using the standard formula \cite{janossy1944}:
\begin{equation}
\label{equation_integral}
N_a = 2 \tau \times N_1 \times N_2.
\end{equation}

In our detector, typical values measured for the single rates at ground level after taking data for $\sim$20 hours are $N_i = 701.57 \pm 0.78 $ Counts Per Minute (CPM). As mentioned before, $\tau = 20 \pm 2 \,\,\mu$s, so that the calculated accidental coincidence rate is
\begin{equation}
N_a = 0.33 \pm 0.03 \,\, \mathrm{CPM}.
\end{equation}

Alternatively, for a direct measurement of the accidental rate the setup also allows a simple disassembly of the upper scintillator detector in such a way that it can be placed sideways, in a geometrical position in which the chance for any particle crossing both detectors is completely negligible (see Fig.~\ref{fig:accidental}). With this setup, the measured accidental rate at ground level after taking data for $\sim$20 hours is
\begin{equation}
N_a = 0.31 \pm 0.02 \,\, \mathrm{CPM}.
\end{equation}
Therefore, both approaches give identical results within uncertainties.

The accidental rate is almost negligible (less than 1\%) when compared with the typical coincidence rate at ground level $N_c$ after $\sim$20 hours running,
\begin{equation}
N_c = 40.00 \pm 0.18 \, \, \mathrm{CPM}.
\end{equation}
For shallow and deep underground sites, this error becomes very relevant as it is comparable or larger than the actual reading (see Section~\ref{sec3}).

\begin{figure}[!hbt]
\centering
\includegraphics[width=0.6\textwidth]{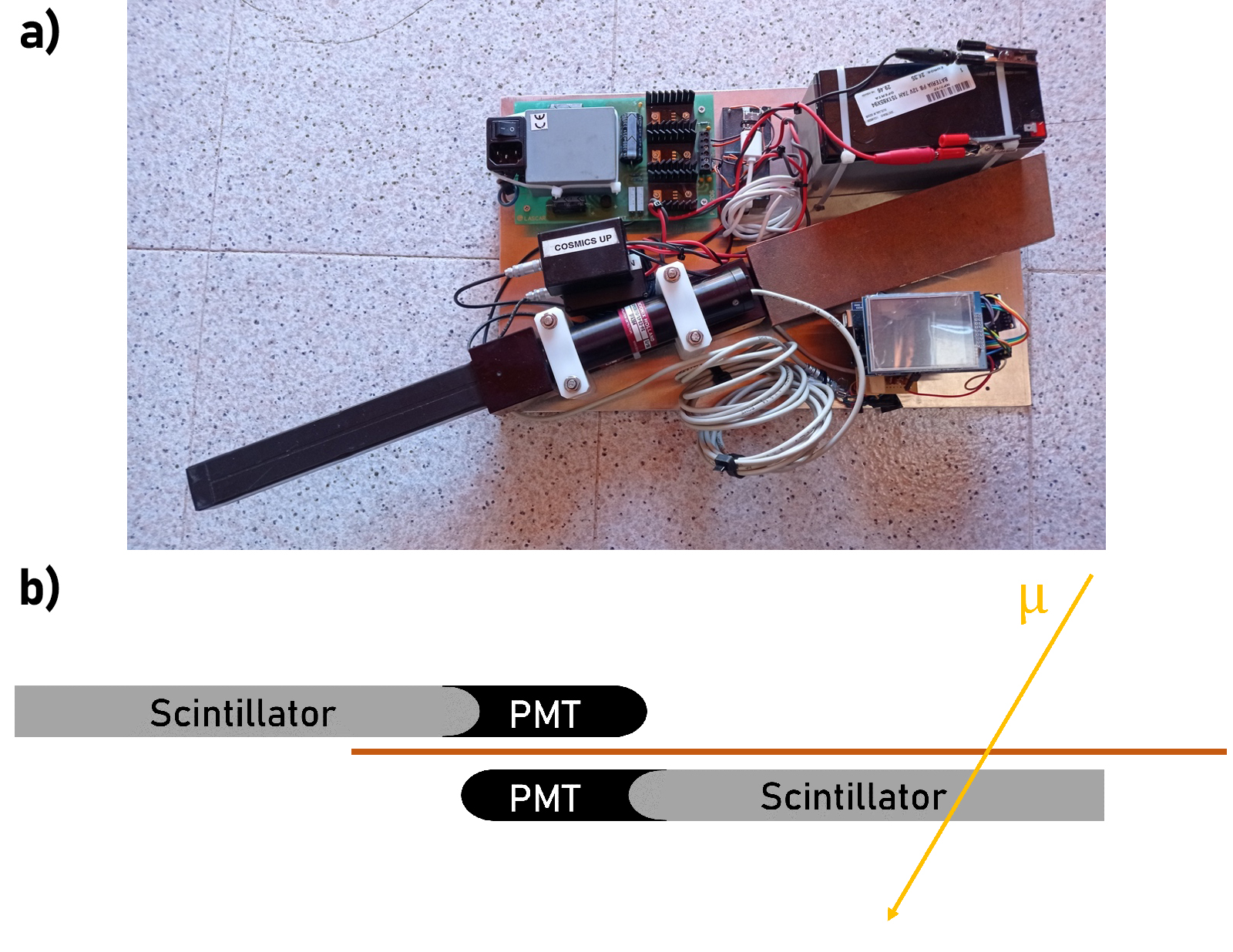}
\caption{a) The muon detector setup for measuring the accidental coincidence rate. The upper scintillator detector set is turned 180 degrees in such a way that the geometrical overlap with the lower scintillator detector set is completely negligible \label{fig:accidental}. b) A schematic of the detector seen from the side. As opposed to Figure~\ref{fig:inside}, now the top scintillator is turned 180 degrees. In this way, only spurious coincidences are counted, since now the muon (in yellow) will only cross one of the two scintillators at a time, meaning it will not yield to a count. }
\end{figure}

{\subsection{Surface measurements}
\label{sec2.4}

After this initial detector characterization, we run longer measurement campaigns in order to minimize statistical errors.

\begin{figure}[!hbt]
\centering
\includegraphics[width=0.6\textwidth]{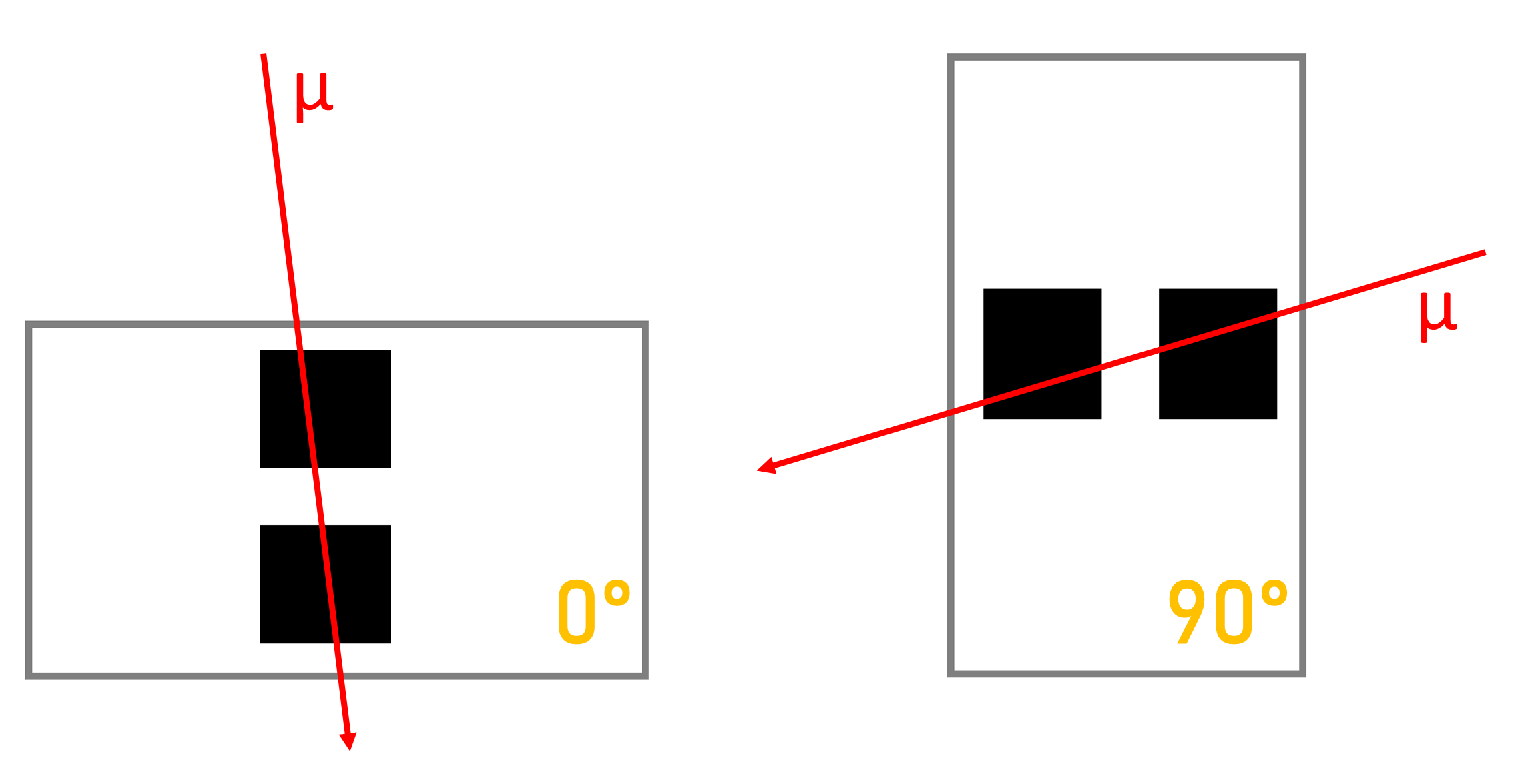}
\caption{Schematics of the suitcase (grey square) containing the scintillators with the PMTs (black squares) for the two different orientations (0 and 90 degrees) considered. On the left, the 0 degrees position which favors the counting of muons coming from a directional orthogonal to the sky (zenith angle of 0 degrees). On the right, the 90 degrees position favoring coincidences induced by muons incoming from shallow angles~\label{fig:valigetta}. }
\end{figure}

In our above-ground laboratory, we measure for $\sim$160 hours with the suitcase placed parallel to the ground (0 degrees). In this case the two detectors are located one above the other, both placed parallel to the ground (see Fig.~\ref{fig:valigetta}), and we obtain:

\begin{equation}
N_H = 41.38 \pm 0.07 \, \, \mathrm{CPM}.
\end{equation}

We then proceed to flip the suitcase 90 degrees, meaning that now the two detectors are on the side of each other (see Fig.~\ref{fig:valigetta}). We then measure for $\sim$327 hours, obtaining:

\begin{equation}
N_V = 8.92 \pm 0.02 \, \, \mathrm{CPM}.
\end{equation}

The stark difference in the counts per minute is due to the strong angular dependence of the incoming muons. In the 90 degrees position, the muons recorded are only the ones reaching the lab at very shallow angles, while in the horizontal position we record most of the incoming muon flux.

\section{Shallow Underground Sites}
\label{sec3}

As explained in Section~$\ref{sec1}$, experiments requiring the lowest possible noise background level locate their sensors in deep underground laboratories around the world, where the rock overburden screens most of the incoming cosmic radiation. Typical depths of underground laboratories are in the 1~km range, obtaining screening factors of cosmic muon flux of $10^5-10^7$ \cite{Trzaska:2019kuk, bellini2012}. Due to their remote location and scarcity, these underground stations do not represent a practical solution for improving the quality of any qubit type affected by cosmic rays. As the lifetime of superconducting qubits limited by ionizing radiation lies already in the millisecond range for transmons \cite{vepsalaainen2020}, it may not be necessary to attain the screening factors achieved in deep underground laboratories and, instead, much shallower sites \cite{PNNL} located in accessible areas may already offer a significant enough screening to bring qubit quality to a level desirable for applications such as error correction and sensing.

\subsection{Shallow Underground Survey}
\label{sec3.1}

We have identified commonly-found locations in city environments providing significant screening factors for cosmic radiation. These are underground stations and side tunnels in underground highways. For the particular case of the city of Barcelona, we have conducted measurements in two such locations using our portable cosmic muon detector described in Section~\ref{sec2}. The first site is the underground station `El Coll-La Teixonera' located under a hill at a depth of approximately 100 meters (see Fig.~\ref{Fig:TMB}).
\begin{figure}[!hbt]
\begin{center}
\includegraphics[width = 0.9\textwidth]{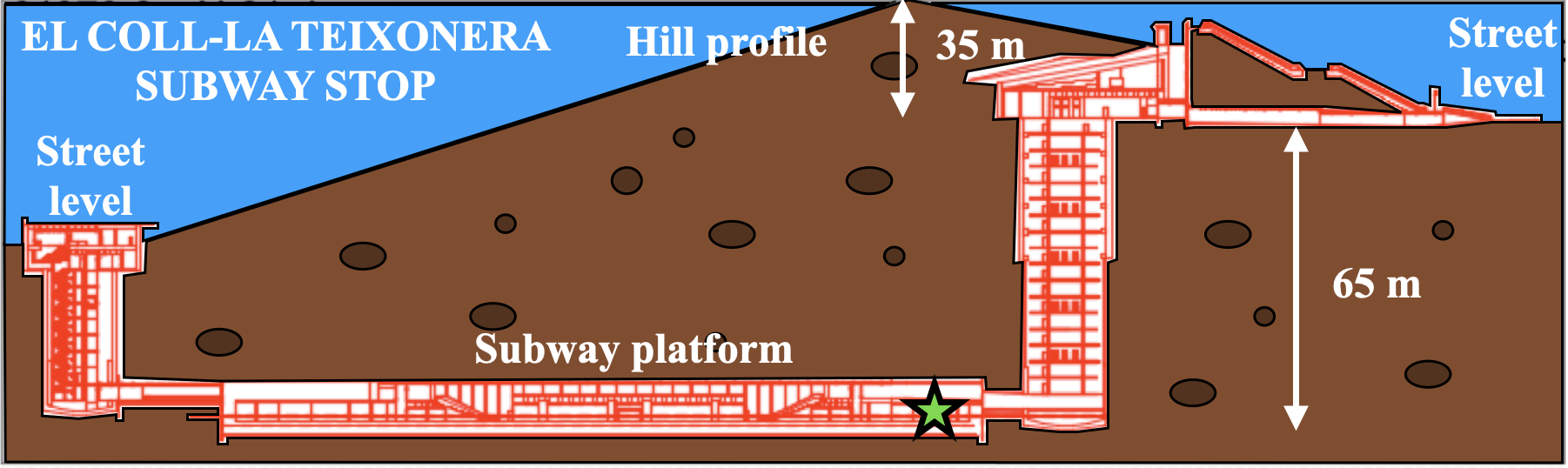}
\caption{\label{Fig:TMB} Profile of the TMB subway station 'El Coll-La Teixonera' where measurements were conducted, indicated by a star icon.}
\end{center}
\end{figure}
The second site is a side tunnel in an underground highway known as the Vallvidrera tunnels running under the `Collserola' hill, with an approximated average depth of 120~m, with a peak height of 170~m (Fig.~\ref{Fig:Vall}).
\begin{figure}[!hbt]
\begin{center}
\includegraphics[width = \textwidth]{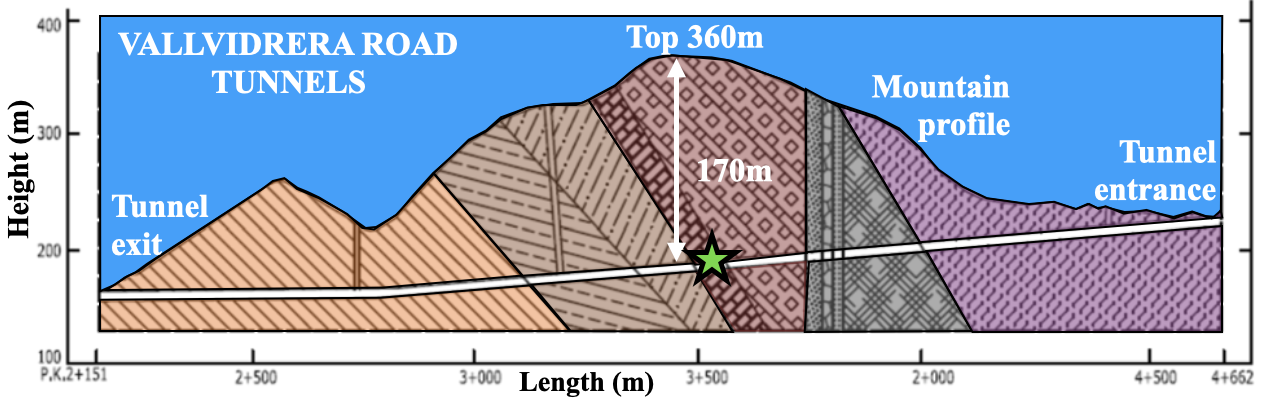}
\caption{\label{Fig:Vall} Profile of the Vallvidrera tunnel under the Collserola hill where measurements were conducted, indicated by a star icon.}
\end{center}
\end{figure}
As a reference of a more realistic depth where qubit laboratories could be easily located, we also performed measurements in a much shallower site 6~m deep in the sub-basement of the ALBA synchrotron located near the Bellaterra campus\footnote{https://www.cells.es/}.

The results are shown in Fig.~\ref{Fig:counts}. All measurements were taken over 48 hours to minimize statistical uncertainty (smaller than the data markers on the plot). Above ground, we measure on average 41.38 CPM. The TMB and Vallvidrera shallow sites investigated display a muon count of 1.4 CPM and 1.1 CPM, respectively, achieving screening factors of 29 and 36 with respect the above ground measurements. The 6~m shallower site achieves a rate of 20 CPM, representing a screening factor of 2. In all data, we have already subtracted the spurious coincidence counts from the apparatus dead time, as detailed in Section~\ref{sec2.3}. In Fig.~\ref{Fig:counts} we show the counts per cm$^{2}$ per minute, calculated from the CPM divided by the effective area of the scintillators in cm$^{2}$.

\begin{figure}[!hbt]
\begin{center}
\includegraphics[width = 0.8\textwidth]{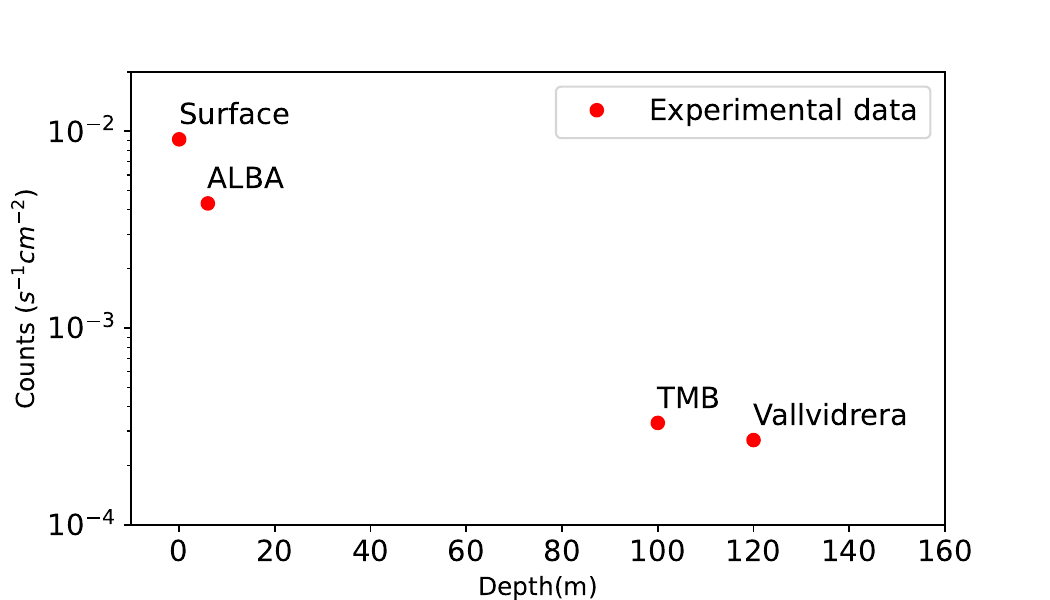}
\caption{\label{Fig:counts} Measurement of muon counts as function of depth. The surface measurement is used as a reference for the two shallow underground sites identified: the Barcelona subway (TMB) and Vallvidrera tunnels. The ALBA synchrotron sub-basement is used as a reference of a typical laboratory location. The statistical uncertainty of all measurements is smaller than the marker size of the plot after averaging over 48h at each location.}
\end{center}
\end{figure}

Inside the Vallvidrera tunnels, we have collected data for single events in our detector with the result of 1047.4 CPM. This represents a factor 1.68 higher than above ground in a usual office space, which is consistent with the presence of heavy minerals in the tunnels generating a higher amount of radioactivity. This is also consistent with measurements in deep underground laboratories. Therefore, it is necessary to equip qubit measurement setups with screening against this type of radioactivity, such as a lead shield. Note that given the low amount of cosmic ray flux in this shallow site, the spallation events discussed in Section~\ref{sec1.1} would also be reduced.

Assuming one is able to screen environmental radioactivity using lead shielding and employ radiopure setups, as is usual in underground sites, the screening factor obtained at 100~m depths would lead to a decrease of events to a level of $\sim3\times10^{-4}$~s$^{-1}$cm$^{-2}$. Using typical device dimensions for superconducting qubit chips of $\sim1$~cm$^{-2}$, a device placed in such a shallow site would experience an average of 0.018 counts per minute, or 0.3~mHz, compared to 0.54 counts per minute, or 9~mHz, above ground.

\subsection{Data cross-check with Geant4 Simulations}
\label{sec3.2}

In order to cross-check the data obtained, we performed Geant4~\cite{GEANT4:2002zbu} simulations of the portable detector above ground and underground, for the two configurations (vertical and horizontal) described in Section~\ref{sec2.4}.

Muon production for the simulation is carried out using the Cosmic-Ray Shower Generator (CRY)~\cite{Hagmann:2007ziw} library. This tool generates cosmic-ray particle shower distributions (muons, neutrons, protons, electrons, photons, and pions) with the proper flux and direction within a user-specified area, elevation and altitude. To optimize computation time only muons that will reach the detector are tracked from the initially generated set.

The simulation is performed for two scenarios, the one in the above-ground laboratory and the one in the shallow underground site of ALBA synchrotron. For the surface simulation, cosmic muons are generated on a square surface at a specific vertical distance from the detector, allowing them to reach the detector with a maximum zenith angle of 70 degrees. According to the cosine square law, this corresponds to simulating 95$\%$ of the incoming flux.

In the underground simulation, a 6-meter-thick volume of standard rock (90$\%$ CaCO$_3$ and 10$\%$ MgCO$_3$, with a density of 2.71~g/cm$^3$) is placed between the detector and the generation plane. To maintain the covered angle at 70 degrees, the dimensions of the generation square need to be increased. However, increasing the generation plane significantly increases computation time. A low-statistics simulation has verified that the results are equivalent to considering a thinner layer of rock with a higher density. Higher-statistics simulations assumed 1 meter of rock with a density of 12~g/cm$^3$.

The rate calculation is based on a reference measurement of 41.38 counts per minute over $\sim$160 hours conducted at the surface. This reference is used to calculate the equivalent real-time of the simulation, which is then used to determine the simulated rate.

The simulation accurately reproduces the flux reduction observed at the ALBA synchrotron location for the 0-degree (horizontal) scenario, see Table~\ref{table:0}. However, there is a discrepancy between the collected data and the simulations for the measurement at 90 degrees (vertical). In this case, only muons with very large azimuthal angles ($\textgreater$~60 degrees) can produce coincidences in the detectors.

Due to geometric limitations, muons with zenith angles greater than 70 degrees are not simulated, which initially would lead to an expected deficit in the simulation compared to the data, with a reduction of approximately 35$\%$. However, an overabundance is observed in the simulated data. A plausible explanation is that muons at larger angles are strongly attenuated by the terrain's topography, leading to a lower count by our detector compared to the simulated muons.

\begin{table}[h!]
\centering
\begin{tabular}{ |p{3cm}|p{3cm}|p{3cm}|p{3cm}|p{3cm}|  }
\hline
\multicolumn{5}{|c|}{Rate (Counts/Minute)} \\

\hline
 & 0$^{\circ}$ - ALBA & 90$^{\circ}$ - ALBA & 0$^{\circ}$ - Surface & 90$^{\circ}$ - Surface \\
\hline
Simulation &19.29~$\pm$~0.62 & 7.17~$\pm$~0.34 &41.38~$\pm$~1.07 & 13.54~$\pm$~0.49\\
Data &20.19~$\pm$~0.07 & 5.89~$\pm$~0.03 &41.38~$\pm$~0.07 & 8.92~$\pm$~0.02\\
\hline
\end{tabular}
\caption{\label{Fig:simulation_res} Comparison between the data and the simulations obtained for the surface and the ALBA synchrotron measurements.}
\label{table:0}
\end{table}

\section{Device Orientation}
\label{sec4}

Since the cosmic ray flux observed at Earth has a strong angular dependence, we investigate in this section the dependence of device orientation in the energy spectrum of the ionizing radiation events impacting the substrate.

\begin{figure}[!hbt]
\begin{center}
\includegraphics[width = 0.8\textwidth]{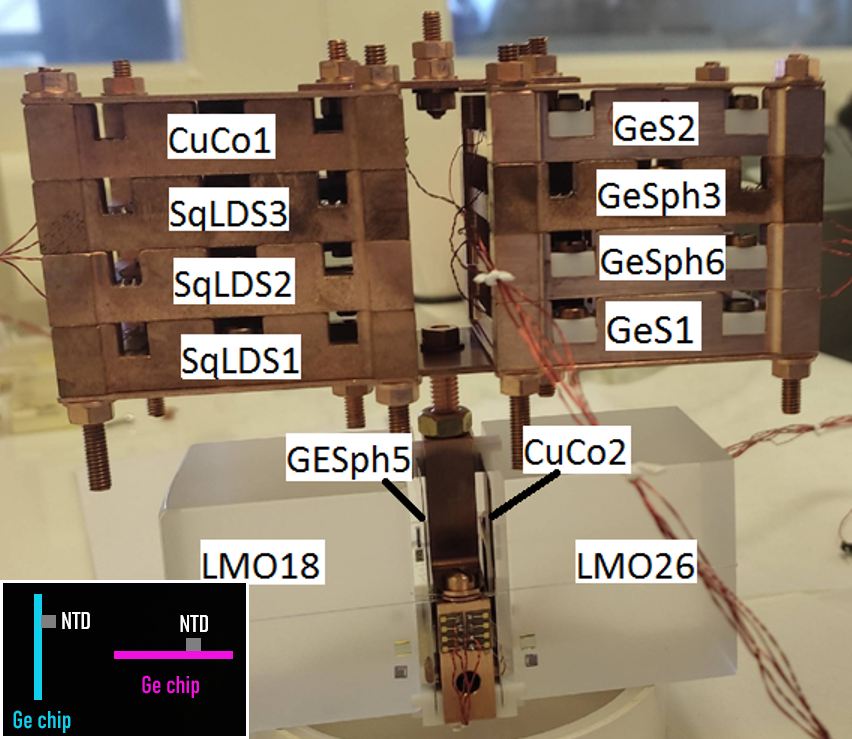}
\caption{\label{Fig:detectors} Detectors mount for the run above ground at the IJCLab in Orsay (France). These detectors were installed and operated at $\sim$15~mK in a dilution refrigerator located above ground. The detector used for the vertical orientation is \textit{CuCo2}, while \textit{GeSph6} is used for the horizontal orientation. In the inset on the bottom left a schematic of the geometry is displayed: in light blue the \textit{vertical} detector, in pink the \textit{horizontal} detector. }
\end{center}
\end{figure}

\subsection{Experimental Methods}

To test the orientation effect of impacts from cosmic radiation, we analyze the data collected by two light detectors developed for the CROSS~\cite{Khalife:2019nyz} and BINGO~\cite{Khalife:2023fmg} experiments.  The detectors were operated simultaneously in a dilution refrigerator located in the above-ground laboratory IJCLab in Orsay (France), for 16.28 hours. For the vertical orientation data, we refer to detector \textit{CuCo2}, while for the horizontal orientation we refer to detector \textit{GeSph6}, see Fig.~\ref{Fig:detectors}. The two detectors consist of a germanium wafer with identical (45$\times$45$\times$0.3)~mm$^3$ dimensions. Each one is equipped with a Neutron-Transportation-Doped (NTD) germanium thermistor~\cite{Wang:1989vk}, a sensor that measures temperature variations induced by particle interactions within the wafer. The NTD ensures high stability and detection efficiency along with high-energy resolution and a low-energy threshold for particle detection. Each NTD is glued directly on top of the germanium wafers using a standardized procedure and they are wire-bonded using a 25~\textmu m gold wire to copper-capton-copper pads thermally connected to the coldest stage of the dilution refrigerator. Hence, the gold wires provide the link between the thermal bath and the NTDs as well as the electrical connections to measure the temperature variations of the detector. In fact, the temperature of the NTD is read out by measuring its voltage
drop while applying a constant bias current through the gold wires.

The energy calibration of the detectors was performed using a mix of features visible in the spectrum of each detector and shining a 600~nm LED placed at room temperature through an optical fiber to the Mixing Chamber of the dilution refrigerator. The LED is pulsed for 2~\textmu s to obtain a characteristic line used to perform a rough energy calibration. This setup is well calibrated~\cite{Auguste:2024xrg}, and as such the peak visible in Fig. 8 is expected to fall at an energy of $\sim$100-300 keV. These fast LED pulses result in a single narrow peak in the energy spectrum, since this bolometer cannot temporally distinguish between an energy deposition caused by a single source or multiple particles at this time scale. As such, all the photons emitted by the LED and subsequently absorbed by the bolometer will appear as a single pulse in the detector.

Multiple methods exist to establish the energy threshold of bolometers~\cite{Mancuso:2017ffg}: we determine the threshold by starting from the distribution of baselines without a distinguishable signal (\textit{noise-only}): we take 1,000 baselines, randomly selected within the collected data, extract for each an amplitude with the same methodology used for particle pulses, and plot these amplitudes. The distribution can be approximated to a Gaussian and can be accurately fit to extract its FWHM. The vertical detector has a FWHM baseline of 0.666~$\pm$~0.010 keV while the horizontal detector has a
FWHM baseline of 0.246~$\pm$~0.003 keV. As such, even if the detectors are equipped with different sensors, we can confidently compare them for energies well above 1~keV, as we expect for cosmic rays impinging these detectors.\\

\begin{figure}[!hbt]
\begin{center}
\includegraphics[width = 0.9\textwidth]{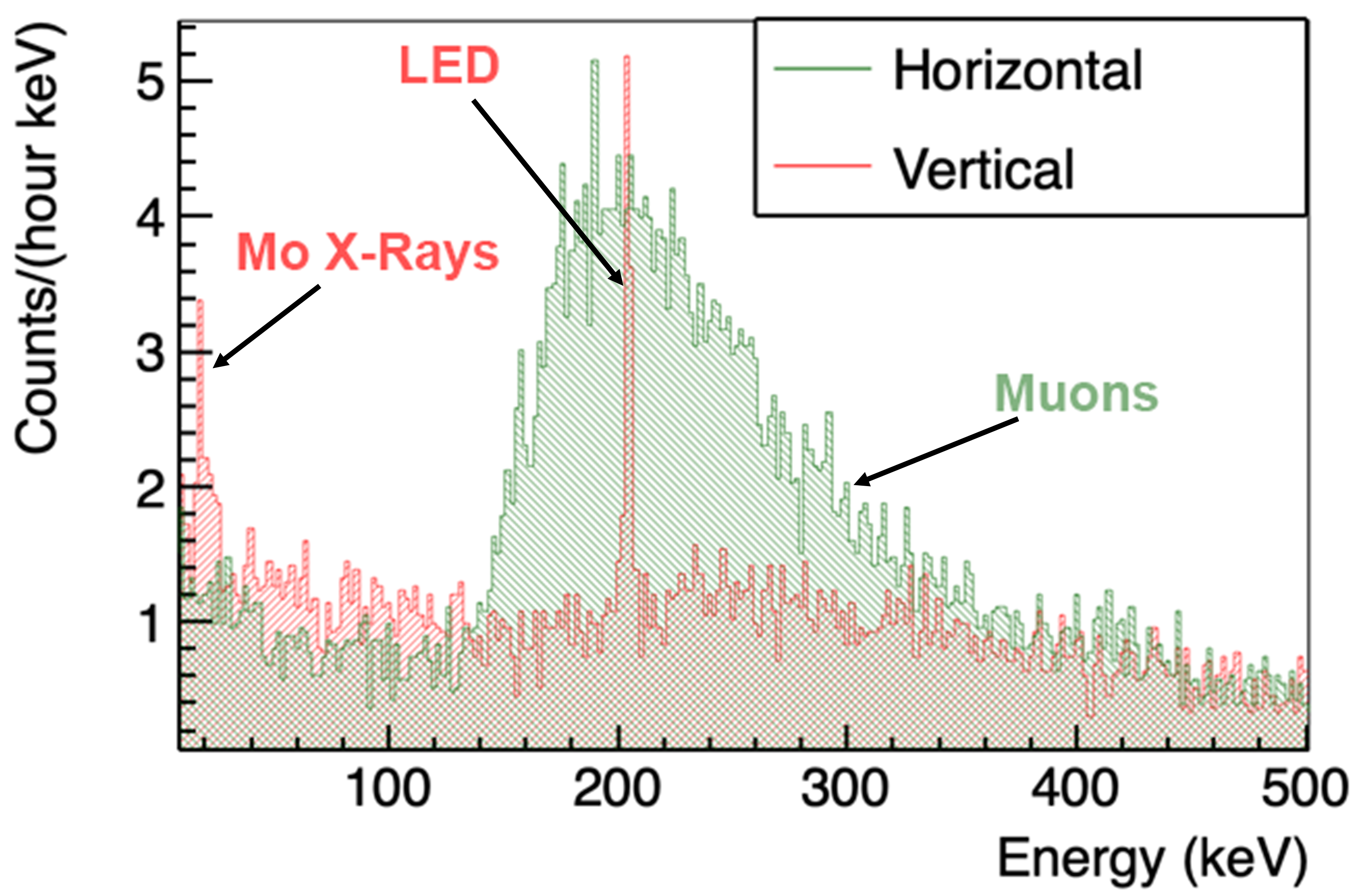}
\caption{\label{Fig:spect} Energy depositions caused by particles in the horizontal (green) and vertical (red) detectors. On the y-axis we plot the counts per hour per keV, on the x-axis the energy of each interaction with a binning of 1~keV. For the horizontal detector there is a noticeable bump between $\sim$140~keV and $\sim$360~keV: we attribute this bump to muon interactions within the sample since it is much less pronounced in the vertical detector. This observation is consistent with simulations of the detectors' response to the cosmic muon flux at this location. The vertical detector in this energy range only exhibits a peak due to the shining of the LED: hence in normal conditions we expect a flat energy spectrum. Furthermore, at low energy, we observe also a peak at $\sim$17 keV due to Mo X-ray fluorescence from the large Li$_2$MoO$_4$ crystal facing the detector (see Fig.~\ref{Fig:detectors}). The position of this peak is used for the energy calibration of the vertical detector.}
\end{center}
\end{figure}

\begin{figure}[!hbt]
\begin{center}
\includegraphics[width = 0.9\textwidth]{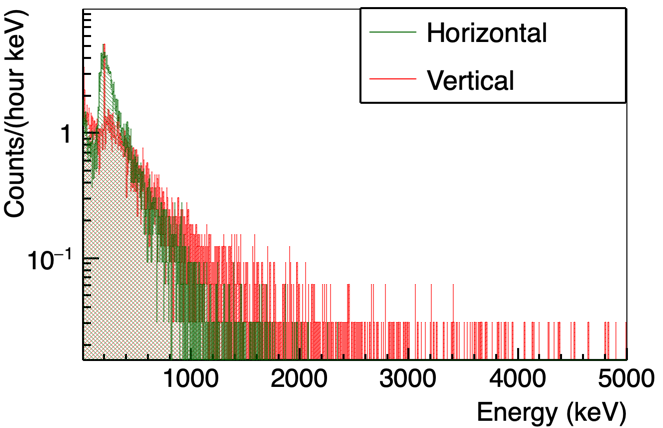}
\caption{\label{Fig:spect2} Energy depositions caused by particles in the horizontal (green) and vertical (red) detectors. On the y-axis we plot the counts per day per keV, on the x-axis the energy of each interaction with a binning of 1~keV. We can still see the muon bump as in Fig.~\ref{Fig:spect}, but we can also notice a longer and more populated tail of highly energetic events ($\gtrsim$~1000~keV) measured by the vertical detector. This is compatible with the different geometry, which allows muons to deposit energy in longer tracks in the detectors placed vertically to the respect of the sky.}
\end{center}
\end{figure}

\subsection{Results and data interpretation}

In Fig.~\ref{Fig:spect}, we plot the energy spectrum up to 500~keV obtained during the simultaneous 16.28 hours run of the vertical and horizontal detectors. We observe a noticeable difference in the shape of both spectra, with a bump in counts in the horizontal detector between $\sim$140~keV and $\sim$360~keV, while the vertical detector has a flat count rate with only a peak at 200~keV due to the shining of the LED through the optical fiber. We decided to show this LED peak, since these are events induced by particles and, as such, are not straightforward to remove from the dataset.

In Fig.~\ref{Fig:spect2}, we plot the complete energy spectra of the two detectors. Here, we see another clear difference, with a longer and more populated tail of highly energetic events ($\gtrsim$~1000~keV) measured by the vertical detector.

\begin{table}[h!]
\centering
\begin{tabular}{ |p{3cm}|p{3cm}|p{3cm}|  }
\hline
\multicolumn{3}{|c|}{Counts/hour} \\
\hline
 & Vertical (no LED) & Horizontal \\
\hline
100-500~keV &5894~$\pm$~77 & 12237~$\pm$~111 \\
$>$100~keV &9058~$\pm$~95 & 14111~$\pm$~119 \\
$>$500~keV & 3164~$\pm$~56 & 1874~$\pm$~43 \\
\hline
\end{tabular}
\caption{Counts per hour recorded by the two detectors for three different energy regions. The excess counts due to the LED peak in the vertical data are subtracted. In the 100-500~keV range the counts in the horizontal detector are more than double, while above 500~keV the vertical detector almost doubles the counts of the horizontal. Overall above 100~keV the vertical detector shows a $\sim$36$\%$ reduction in total counts.}
\label{table:1}
\end{table}

Overall, in Table~\ref{table:1} we report the counts per hour recorded by the two detectors for 3 different energy regions, in order to highlight the differences induced by the different geometries. Once the excess counts due to the LED peak in the vertical data are subtracted, in the 100-500~keV range we notice that the counts in the horizontal detector are more than double the ones recorded by the vertical detector. Above 1000~keV, the situation is reversed, with the vertical detector almost doubling the counts of the horizontal detector. We do not focus on events at lower energies for two reasons: cosmic rays release a larger amount of energy compared to other ionizing particles and in bolometers most of the counts at low energies are due to exponential rises close to their energy threshold, a feature that is detector-specific and still not fully explained~\cite{Fuss:2022fxe}. Hence, the choice of 100~keV threshold is arbitrary, but safe as it encompasses the physical phenomena we want to study with this setup. Overall, above 100~keV the vertical detector shows a $\sim$36$\%$ reduction in total counts.\\
Muons with energies on the order of 1-100 GeV cross materials along a rather linear track and release energy at a constant rate. Hence, the energy spectra we are presenting are quite straightforward to interpret: in the vertical mounting, due to the angular dependence of the muon flux, a lower overall count rate is expected, since there is less available surface orthogonal to the flux. However, the horizontal position ensures that, statistically, most of the muons track will be short and in a confined energy range, hence the observed muon bump. The vertical geometry allows for longer tracks and more energy release for cosmic rays coming at orthogonal angles with respect to the ground, or, equivalently, at shallow zenith angles.

Compared to a typical present-day qubit chip~\cite{48651}, which has $\sim$(10$\times$10$\times$0.3)~mm$^3$ dimensions, these germanium wafers are significantly larger. Hence, the effects due to the mounting geometry are clearer and enhanced compared to smaller chips. However, in the quest to build larger and larger quantum processors, the mounting orientation, as we are showing with our results, might start to play a role in laboratories that sustain a large muon flux. Additionally, the wafers are composed by germanium, instead of silicon or sapphire, as it is typical in quantum processors. Germanium has more than double the density of silicon and a higher density than sapphire, so we would expect a lower count rate on a comparable quantum computing chip geometry.

As a final remark, the mounting orientation of the chip can have a significant effect in mitigating the impact of ionizing radiation on quantum processors. However, it is unclear at present if it would be preferable to have a lower count rate, but with more disruptive energy depositions, like in the vertical mounting scenario, or if it is better to minimize high energy releases, but endure more frequent particle interactions with the chip as in the horizontal mounting.

\section{Conclusions}
In this work, we have studied two methods to reduce the flux of cosmic muons compatible with experiments involving superconducting qubits. \\
The first method consists in locating qubit experiments in shallow underground sites, which could be as low as 10~m below surface. This small depth already provides strong protection against neutrons, and it reduces the cosmic muon flux by a factor 2. At deeper sites at 100~m depth we observe a reduction of cosmic muon flux with factors on the order of 30. Notably, the shallow sites identified in this work are located in urban areas with far greater accessibility than remote, deep underground laboratories, providing potential sites in which to install future, large-scale quantum processors. We have validated this method using a home-built, portable cosmic muon counter consisting of two plastic scintillators integrated in a coincidence counting electronics circuit. The data collected by this portable detector has been cross-checked through Geant4 simulations. \\
The second method is linked with the mounting orientation of the chip inside the dilution refrigerator. We employ two identical germanium wafers equipped with a particle sensor in a laboratory above ground to show how the vertical orientation of the chip reduces the count rate by 36$\%$ above 100~keV. However, this comes at the expense of enduring a higher count rate at large energy depositions within the substrate. In the present day, since the energy range that is most detrimental to qubit performances has not yet been determined, it is unclear which mounting orientation should be preferred.

The unquestionable evidence of the effect of cosmic muons on qubits requires addressing this source of noise and instability with as much priority as the rest of sources of noise, such as TLS surface defects. Decreasing the impact of cosmic muons will only be possible through a combination of techniques to reduce their flux, following methods similar to the ones proposed in this work, and their impact with on-chip mitigation techniques, such as phonon and quasiparticle suppression. Understanding the physics of cosmic muons interacting with qubit devices is an important future step to further suppress their effect and lead to higher quality qubits to build more robust quantum computers and more stable quantum sensors.

\label{conclusions}

\section*{Acknowledgments} We would like to thank Marcos Gonz\'alez and Adela Rubio from `Transport Metropolitans de Barcelona' (TMB), Jordi Quevedo from `T\'unels de Barcelona i Cad\'i' and Joan Casas and Ramon Pascual from the ALBA synchrotron for granting us access to their premises to conduct the cosmic muon analysis. We also thank Carlos Peña-Garay at LSC and Boris Nedyalkov for fruitful discussions. We acknowledge funding from the Ministry of Economy and Competitiveness and Agencia Estatal de Investigación (RYC2019-028482-I, PCI2019-111838-2, PID2021-122140NB-C31), and the European Commission (FET-Open AVaQus GA 899561, QuantERA) and program `Doctorat Industrial' of the Agency for Management of University and Research Grants (2020 DI 41; 2020 DI 42). IFAE is partially funded by the CERCA program of the Generalitat de Catalunya. E.~B. is supported by the Grant FJC2021-046443-I funded by MCIN/AEI/10.13039/501100011033. This study was supported by MICIIN with funding from European Union NextGenerationEU(PRTR-C17.I1) and by Generalitat de Catalunya.
Finally, we would like to thank the CROSS and BINGO collaborations for providing us with data collected with their Ge-based bolometric detectors.

\end{document}